\documentstyle{mn}
\newif\ifAMStwofonts
\def\ee #1 {\times 10^{#1}}
\def\ut #1 #2 { \, \rmn{#1}^{#2}}
\def\u #1 { \, \rmn{#1}}

\def\percc {\,\rmn{cm}^{-3}}

\def\half{{\textstyle \frac{1}{2}}}

\let\grad=\nabla
\def\cross{\bmath{\times}}
\def\curl {\grad \bmath{\times}}
\def\div {\grad \cdot}

\def\E{\bmath{E}}            % E
\def\Epa{\bmath{E'_\parallel}}  % E'_||
\def\Epe{\bmath{E'_\perp}}  % E'_perp
\def\J{\bmath{J}}            % J
\def\B{\bmath{B}}            % B
\def\v{\bmath{v}}            % v
\def\k{\bmath{k}}            % k
\def\Bh{\bmath{\hat{B}}}     % B
\def\fh{\bmath{\hat{\phi}}}  % phi
\def\rh{\bmath{\hat{r}}}     % r
\def\zh{\bmath{\hat{z}}}     % z
\def\vf{v_{\phi}}            % v_phi
\def\vk{v_{K}}               % v_K

\def\dvp{\bmath{\delta\v_{\perp}}}
\def\dEp{\bmath{\delta\E'_{\perp}}}
\def\dBp{\bmath{\delta\B_{\perp}}}
\def\dJp{\bmath{\delta\J_{\perp}}}

\newcommand{\delt} [1] {\frac{\partial #1}{\partial t}}

\newcommand{\delf} [1] {\frac{\partial #1}{\partial \phi}}

\ifoldfss
  \newcommand{\rmn}[1] {{\rm #1}}

  \ifCUPmtlplainloaded \else
    \NewTextAlphabet{textbfit} {cmbxti10} {}
    \NewTextAlphabet{textbfss} {cmssbx10} {}
    \NewMathAlphabet{mathbfit} {cmbxti10} {} % for math mode
    \NewMathAlphabet{mathbfss} {cmssbx10} {} %  "   "    "
  \fi
  \ifAMStwofonts
    \ifCUPmtlplainloaded \else
      \NewSymbolFont{upmath} {eurm10}
      \NewSymbolFont{AMSa} {msam10}
      \NewMathSymbol{\upi}     {0}{upmath}{19}
      \NewMathSymbol{\umu}     {0}{upmath}{16}
      \NewMathSymbol{\upartial}{0}{upmath}{40}
      \NewMathSymbol{\leqslant}{3}{AMSa}{36}
      \NewMathSymbol{\geqslant}{3}{AMSa}{3E}

      \let\leq=\leqslant 
       
    \fi
  \fi
\fi % End of OFSS
\ifnfssone
  \newmathalphabet{\mathit}
  \addtoversion{normal}{\mathit}{cmr}{m}{it}
  \addtoversion{bold}{\mathit}{cmr}{bx}{it}
  \newcommand{\rmn}[1] {\mathrm{#1}}

  \newmathalphabet{\mathbfit} % math mode version of \textbfit{..}
  \addtoversion{normal}{\mathbfit}{cmr}{bx}{it}
  \addtoversion{bold}{\mathbfit}{cmr}{bx}{it}
  \newmathalphabet{\mathbfss} % math mode version of \textbfss{..}
  \addtoversion{normal}{\mathbfss}{cmss}{bx}{n}
  \addtoversion{bold}{\mathbfss}{cmss}{bx}{n}
  \ifAMStwofonts
    \ifCUPmtlplainloaded \else
      \UseAMStwoboldmath
      \makeatletter
      \new@mathgroup\upmath@group
      \define@mathgroup\mv@normal\upmath@group{eur}{m}{n}
      \define@mathgroup\mv@bold\upmath@group{eur}{b}{n}
      \edef\UPM{\hexnumber\upmath@group}
      \new@mathgroup\amsa@group
      \define@mathgroup\mv@normal\amsa@group{msa}{m}{n}
      \define@mathgroup\mv@bold\amsa@group{msa}{m}{n}
      \edef\AMSa{\hexnumber\amsa@group}
      \makeatother
      \mathchardef\upi="0\UPM19
      \mathchardef\umu="0\UPM16
      \mathchardef\upartial="0\UPM40
      \mathchardef\leqslant="3\AMSa36
      \mathchardef\geqslant="3\AMSa3E

      \let\leq=\leqslant 

    \fi
  \fi
\fi % End of NFSS release 1
\ifnfsstwo
  \newcommand{\rmn}[1] {\mathrm{#1}}

  \DeclareMathAlphabet{\mathbfit}{OT1}{cmr}{bx}{it}
  \SetMathAlphabet\mathbfit{bold}{OT1}{cmr}{bx}{it}
  \DeclareMathAlphabet{\mathbfss}{OT1}{cmss}{bx}{n}
  \SetMathAlphabet\mathbfss{bold}{OT1}{cmss}{bx}{n}
  \ifAMStwofonts
    \ifCUPmtlplainloaded \else
      \DeclareSymbolFont{UPM}{U}{eur}{m}{n}
      \SetSymbolFont{UPM}{bold}{U}{eur}{b}{n}
      \DeclareSymbolFont{AMSa}{U}{msa}{m}{n}
      \DeclareMathSymbol{\upi}{0}{UPM}{"19}
      \DeclareMathSymbol{\umu}{0}{UPM}{"16}
      \DeclareMathSymbol{\upartial}{0}{UPM}{"40}
      \DeclareMathSymbol{\leqslant}{3}{AMSa}{"36}
      \DeclareMathSymbol{\geqslant}{3}{AMSa}{"3E}

      \let\leq=\leqslant 

    \fi
  \fi
\fi % End of NFSS release 2
\ifCUPmtlplainloaded \else
  \ifAMStwofonts \else % If no AMS fonts
    \def\upi{\pi}
    \def\umu{\mu}
    \def\upartial{\partial}
  \fi
\fi
\input epsf
\title[Balbus-Hawley instability in weakly ionized discs]
{The Balbus-Hawley instability in weakly ionized discs}
\author[Mark Wardle]
       {Mark Wardle\\
Special Research Centre for Theoretical Astrophysics, University
of Sydney, NSW 2006, Australia }
\date{1999 February 26}
\pagerange{\pageref{firstpage}--\pageref{lastpage}}
\pubyear{1999}
\begin{document}
\maketitle
\label{firstpage}
\begin{abstract}
MHD in protostellar discs is modified by the Hall current when the 
ambipolar diffusion approximation breaks down.  Here I examine the 
Balbus-Hawley (magnetorotational) instability of a weak, vertical 
magnetic field within a weakly-ionized disc.  Vertical stratification 
is neglected, and a linear analysis is undertaken for the case that 
the wave vector of the perturbation is parallel to the magnetic field.

The growth rate depends on whether the initial magnetic field is 
parallel or antiparallel to the angular momentum of the disc.  The 
parallel case is less (more) unstable than the antiparallel case if 
the Hall current is dominated by negative (positive) species.  The 
less-unstable orientation is stable for $\chi\la 0.5$, where $\chi$ is 
the ratio of a generalised neutral-ion collision frequency to the 
Keplerian frequency.  The other orientation has a formal growth rate 
of order the Keplerian angular frequency even in the limit 
$\chi\rightarrow 0 $!  In this limit the wavelength of the fastest 
growing mode tends to infinity, so the minimum level of ionization for 
instability is determined by the requirement that a wavelength fit 
within a disc scale height.  In the ambipolar diffusion case, this 
requires $ \chi > v_A/c_s $; in the Hall case this imposes a 
potentially much weaker limit, $\chi > v_A^2 / c_s^2$.
\end{abstract}
\begin{keywords}
accretion discs --
instabilities --
magnetic fields --
MHD --
stars: formation
\end{keywords}

\section{Introduction}
The Balbus-Hawley (magnetorotational) instability (Velikhov 1959; 
Chandrasekhar 1961) of weak magnetic fields in accretion discs drives 
MHD turbulence that transports angular momentum radially outwards 
(Balbus \& Hawley 1991, Hawley \& Balbus 1991; Stone et al 1996), and 
is thought to play an important role in the evolution and dynamics of 
astrophysical accretion discs.  The instability has also been invoked 
as a component of a disc dynamo model, in which the instability 
creates radial field from vertical field, the shear in the disc 
creates azimuthal field from the radial component, and the Parker 
instability creates vertical from azimuthal field and expels flux from 
the disc (Tout \& Pringle 1992).

If the disc is sufficiently ionized that flux-freezing is a good 
approximation, the characteristic growth rate and vertical wave number 
of the instability are of order $ \Omega $ and $ \Omega/v_A $ 
respectively, where $ \Omega $ and $ v_A $ are the Keplerian angular 
frequency and the Alfv\'en speed in the disc.  The growth of the 
instability in a weakly-ionized, magnetized disc is of interest for 
theories of star formation and the subsequent evolution of 
protostellar discs.  In this context, the degree of coupling between 
the field and neutral gas is determined by the trace charged species 
that are produced by cosmic-ray, radioactive or thermal ionization 
(Hayashi 1981) or the X-ray flux from the central, magnetically active 
star (Glassgold, Najita \& Igea 1997).  The level of ionization may 
only be sufficient to couple a magnetic field to the material in the 
surface regions, where incident X-rays and cosmic rays are absorbed, 
restricting magnetic activity to the surface layers over 
a large portion of the disc (Gammie 1996; Wardle 1997).

Two models for field diffusion are generally adopted when considering 
the breakdown of flux freezing in weakly-ionized gas.  At relatively 
low densities, the magnetic field can be regarded as frozen into the 
ionized component of the gas and drifts with it through the neutrals, 
a process referred to as ambipolar diffusion (Spitzer 1978).  The 
linear growth of the instability has been examined in the 
presence of ambipolar diffusion (Blaes \& Balbus 1994), and the 
nonlinear development has been investigated (MacLow et al 1995; Hawley 
\& Stone 1998).  The instability grows only if the neutral-ion 
coupling time scale is less than about $\Omega^{-1}$.  At high 
densities collisions with neutral particles stop the charged species 
drifting with the magnetic field, and the fluid can be treated as 
resistive -- this is the limit of Ohmic diffusion.  The 
behaviour of the instability in the presence of Ohmic diffusion has 
also been considered in the linear (Jin 1996) and nonlinear regimes 
(Sano, Inutsuka \& Miyama 1998).  In this case it is the resistive 
diffusion time scale that has to be compared with $\Omega^{-1}$.
 
However, both ambipolar and resistive diffusion are generally poor 
approximations in protostellar discs, where the gas density is 
sufficient for collisions with neutrals to partially or completely 
decouple ions and grains from the magnetic field, but electrons are 
either well-coupled or partially coupled to the field (Wardle \& 
K\"onigl 1993).  Instead, the gas is in the Hall regime, where 
variations in the decoupling amongst the charged species produces an 
overall handedness in the fluid with respect to the magnetic field 
(Wardle \& Ng 1999).  That is, the fluid dynamics is no longer 
invariant under a global reversal of the magnetic field.  For example, 
left- and right-circularly polarized Alfv\'en waves propagate at 
different speeds, and damp at different rates (Pilipp, Hartquist \& 
Havnes 1987; Wardle \& Ng 1999).  This qualitative change in the wave 
modes supported by the fluid implies that the dynamical evolution of 
the magnetized gas will be dramatically modified.

This paper examines the modifications to the Balbus-Hawley instability 
in weakly-ionized gas.  In Section 2 I write down the fluid equations 
appropriate for a weakly-ionized, magnetized, near-Keplerian disc.  
The linearized equations and dispersion relation are obtained in 
Section 3 for pertubations with wave vector parallel to an initially 
vertical field, neglecting stratification of the disc.  Two important 
dimensionless parameters emerge from this analysis: $\chi$, the ratio 
of the neutral coupling frequency to the Keplerian frequency, and the 
ratio of the Hall and Pedersen conductivities, $\sigma_1$ and 
$\sigma_2$.  The dependence of the growth rate on these parameters is 
presented in Section 4, where I show that the growth rate of the 
instability in the Hall regime is dependent on whether the initial 
vertical field is parallel or anti-parallel to the disc angular 
velocity vector.  In particular, the surprising result emerges that 
the growth rate may be non-zero even in the limit $\chi \rightarrow 
0$.  This result is interpreted physically using a two-fluid model of 
the Hall limit in Section 5.  The implications for the evolution and 
dynamics of protostellar discs are discussed in Section 6, and the 
paper is summarized in Section 7.

\section{Formulation}
\label{sec:formulation}

I begin by writing down the equations describing the fluid dynamics of 
a weakly-ionized, magnetized, non-self-gravitating disc moving in the 
potential of a central point mass $ M $.  The definition of 
weakly-ionized in this context is that the inertia of the charged 
species can be neglected.  This constrains the frequencies of interest 
to be below the collision frequency of each charged species with the 
neutrals.  Under this assumption the charged species can be 
incorporated into a conductivity tensor $ \bsigma $, and seperate 
equations of motion are not needed.

Following standard practice, I write the equations in a local 
Keplerian frame, so $ \v $ is the \emph{departure} from Keplerian 
motion -- the fluid velocity in the laboratory frame is $ \v+\v_K $, 
where the Keplerian velocity $ \v_K = \sqrt{GM/r} \, \fh $ in the 
canonical cylindrical coordinate system $ (r,\phi,z) $.  The partial 
time derivatives in the Keplerian frame, written as $ 
\partial/\partial t $ in the equations to follow, correspond to 
$\delt{} + \Omega\delf{}$ in the laboratory frame.
The continuity equation is
\begin{equation}
\delt{\rho} + \div(\rho \v) = 0 \,.	
	\label{eq:continuity}
\end{equation}
Near the disc midplane, on scales small compared to the disc 
thickness, the radial gradient in the gravitational potential is 
cancelled by the portion of the centripetal term associated with exact 
Keplerian motion.  Tidal effects can be neglected and the momentum 
equation becomes
\begin{equation}
% \[
\delt{\v} + (\v \cdot \grad)\v +
\frac{1}{\rho} \grad P - 2\Omega\vf \rh + \half \Omega v_r\fh
% \]
     = \frac{\J\cross\B}{c\rho} \,,	
	\label{eq:momentum}
\end{equation}
where $ \Omega = \vk/r $ is the Keplerian frequency.  The pressure 
gradient term does not appear in the linearized equations, which are 
restricted to Alfv\'en modes, thus we need not consider an equation of 
state.  The current density 
is given by  Amp\'ere's law
\begin{equation}
\J = \frac{c}{4\pi}\grad\cross \B \,.
\label{eq:j_curlB}
\end{equation}
The evolution of the magnetic field $ \B $ is determined by the 
induction equation
\begin{equation}
\delt{\B} = \curl (\v \cross \B) - c \curl \E' - 
{\textstyle \frac{3}{2}} \Omega B_r \fh  \,,
\label{eq:induction}
\end{equation}
where $ \E' $
is the electric field in the frame comoving with the neutrals, 
and $ \B $ must satisfy
\begin{equation}
\div \B = 0 \,.
\label{eq:divB}
\end{equation}
The last term in eq. (\ref{eq:induction}) represents the generation of 
toroidal field from the radial component by Keplerian shear.

In the weakly-ionized limit relevant to protostellar discs, $ \J $ and 
$ \E' $ are related by a conductivity tensor, $ \bsigma $, that is 
determined by the abundances and drifts of the charged species through 
the neutral gas.  I characterize each charged species $j$ by particle 
mass $m_j$ and charge $Z_je$, number density $n_j$, and drift velocity 
through the neutral gas $\v_j$, and assume overall charge neutrality, 
$\sum_j n_j Z_j = 0$.  Assume that the fluid evolves on a time scale 
that is long compared to the collision time scale of any type of 
charged particle with the neutrals.  Then each charged particle drifts 
through the neutrals at a rate and direction determined by the 
instantaneous Lorentz force on the particle and the drag force contributed 
by neutral collisions:
\begin{equation}
Z_j e\left(\E' + {\v_j \over c} \cross \B\right) - \gamma_j m_j \rho \v_j
= 0 \,,
\label{eq:charged_drift}
\end{equation}
where $ \gamma_j = <\sigma v>_j / (m+m_j) $, $ <\sigma v>_j $ is the rate coefficient 
for momentum transfer by collisions with the neutrals and $ m $ is the 
mean neutral particle mass.  The relative importance of the Lorentz 
and drag forces in balancing the electric force on species $j$
is determined by the Hall parameter,
\begin{equation}
\beta_j= \frac{Z_jeB }{ m_j c} \, \frac{1 }{ \gamma_j \rho}\,
\label{eq:Hall_parameter}
\end{equation}
which is the product of the gyrofrequency and the time scale for 
momentum exchange by collisions with the neutrals.  This parameter 
controls the magnitude and direction of the charged particle drift and 
whether a particular species may be regarded as tied to the magnetic 
field by electromagnetic stresses ($|\beta_j| \gg 1$ implies $c\E' 
\approx -\v_j \cross \B$), to the neutral gas by collisions 
($|\beta_j|\ll 1$ implies $ \gamma_j m_j \rho\v_j \approx Z_j e\E'$), 
or to neither ($|\beta_j| \sim 1$).

The current density can be expressed in terms of $\E'$ and $\B$ by inverting
eq (\ref{eq:charged_drift}) for 
$ \v_j $, and forming $\J= e\sum_j n_j Z_j \v_j$.  The resultant 
expression can be written as
\begin{equation}
	\J = \sigma_{\parallel} \Epa + \sigma_1 \Bh \cross \Epe + 
	\sigma_2 \Epe
	\label{eq:jsigmaE}
\end{equation}
where $ \Epa $ and $ \Epe $ are the decomposition of $ \E' $ 
into vectors parallel and perpendicular to $ \B $ respectively and
the components of $ \bsigma $ are the 
conductivity parallel to the magnetic field,
\begin{equation}
	\sigma_{\parallel} =  \sum_j \frac{n_jZ_j^2e^2}{m_j 
	\gamma_j\rho} = \frac{ec}{B}\sum_{j} n_j Z_j \beta_j \,,
	\label{eq:sigma0}
\end{equation}
the Hall conductivity,
\begin{equation}
	\sigma_1 = \frac{ec}{B}\sum_{j}\frac{n_j Z_j}{1+\beta_j^2}\,,
	\label{eq:sigma1}
\end{equation}
and the Pedersen conductivity
\begin{equation}
	\sigma_2 = \sum_j \frac{n_jZ_j^2e^2}{m_j 
	\gamma_j\rho}\, \frac{1}{1+\beta_j^2} = \frac{ec}{B}\sum_{j}\frac{n_j Z_j \beta_j}{1+\beta_j^2}
	\label{eq:sigma2}
\end{equation}
(Cowling 1957; Norman \& Heyvaerts 1985; Nakano \& Umebayashi 
1986).  Later I shall refer to the total conductivity 
perpendicular to the field,
\begin{equation}
	\sigma_{\perp} = \sqrt{\sigma_1^2 + \sigma_2^2} \,.
	\label{eq:sigma_perp}
\end{equation}
 
The second forms given for the parallel and Pedersen components 
in eqs (\ref{eq:sigma0}) and (\ref{eq:sigma2}) are 
useful for comparing their magnitudes with the Hall component.
$\sigma_{\parallel}$ and $\sigma_2$ are always positive, whereas 
$\sigma_1$ may take on either sign, depending how the magnitudes of 
the Hall parameters of different charged species are distributed.  The 
ambipolar diffusion limit is recovered when $|\beta_j| \gg 1$ for most 
species, when $\sigma_{\parallel}\gg\sigma_2\gg|\sigma_1|$.  The 
Ohmic limit is approached when $|\beta_j| \ll 1$ for most species, 
implying that $\sigma_2 \approx \sigma_{\parallel} \gg |\sigma_1|$.  
Less well-known is the Hall limit, which can be approached when the 
magnitudes of the Hall parameters associated with particles of one 
sign of charge are large, and those of the other sign are small.  Then 
$|\sigma_1|$ is greater than both $\sigma_{\parallel}$ and $\sigma_2$ 
(see Section \ref{sec:two-fluid}).

Numerical evaluation of the Hall parameters for ions and electrons 
(e.g. Wardle \& Ng 1999) gives $\beta_i \approx 0.46 
B_{\mathrm{mG}}n_{10}^{-1}$ and $\beta_e \approx -350 n_{10}^{-1} B_{\mathrm{mG}} 
T_{100}^{-1/2}$, where $B_{\mathrm{mG}}$ is the magnetic field 
strength in mG, $n_{10}$ and $T_{\mathrm{100}}$ are the density of 
hydrogen nuclei and electron temperature normalised by $10^{10}\percc 
$ and $100 \u K $ respectively.  For all but the smallest grains, one 
finds $|\beta_g| \ll 1 $, on account of their large geometric 
cross-section.  The role played by grains is uncertain as they rapidly 
settle to the disc midplane.  As the gas density at a particular 
cylindrical radius varies by many orders of magnitude from the disc 
midplane to the surface, and the magnetic field strength may range 
from the milliGauss strengths found in the parent cloud to values a 
hundred times greater, one expects large regions of protostellar discs 
to be in a regime intermediate between the ambipolar diffusion and 
resistive limits, where $\sigma_1$ is non-negligible (Wardle \& Ng 1999).

\section{Linearization}
\label{subsec:linearization}

The equations are linearized about an initial steady state in which $ 
\v$, $ \grad P $, $ \E' $ and $ \J $ vanish, and $ \B = B \zh $.  This 
is appropriate for a box that is small in all dimensions compared to 
the scale height.  Changes in the conductivity associated with the 
perturbations do not appear in the linearized equations, as $ \E' $ 
vanishes in the unperturbed state.  Thus I need not specify how the 
charged particle abundances respond to the perturbations -- all that 
is needed are the components of the conductivity tensor in the 
unperturbed state.

Assuming solutions to the linearized equations of the form 
$ \exp i(\omega t - k z) $, the linear system splits into one subsystem 
corresponding to sound waves propagating along the magnetic field, 
and the interesting subsystem, for which the $ z $-components of 
the perturbed velocity, magnetic field, current density and electric 
field all vanish.  Denoting the resulting two-dimensional vectors by the 
subscript $ \perp $, one finds the following expressions for the 
perturbations.

Amp\'ere's law allows $ \dJp $ to be expressed as
\begin{equation}
	\dJp = \frac{ick}{4\pi}\, \left(
	\begin{array}{cc}
		0 & -1  \\
		1 & 0
	\end{array}
	\right) \dBp
	\label{eq:deltaJperp}
\end{equation}
where the matrix on the right hand side represents the cross product 
operator
$\zh\cross $.  Substituting this into the linearized momentum 
equation allows the perturbations in $ \v $ and $ \B $ to be 
related:
\begin{equation}
	\left(
	\begin{array}{cc}
		\omega & 2i\Omega  \\
		-\half i\Omega & \omega
	\end{array}
	\right) \frac{\dvp}{v_A} = kv_A\frac{\dBp}{B}\,.
	\label{eq:deltavperp}
\end{equation}
In the absence of rotation ($ \Omega = 0 $), this reduces to the 
standard relationship for Alfv\'en waves, $ \dBp/B = (\omega/kv_A) 
\dvp/v_A $. The ratio $ \omega/kv_A $ is 1 in the ideal case, but 
is modified in the non-ideal case (Wardle \& Ng 1999).
The off-diagonal terms in the matrix on the left hand side are 
introduced by the Coriolis and centripetal terms in eq. 
(\ref{eq:momentum}), implying that the unstable mode is a modified 
Alfv\'en mode.

The linearized induction equation becomes, after substituting for $ 
\dvp $, 
\[
	\left(
	\begin{array}{cc}
		\omega^2 - k^2 v_A^2 + 3\Omega^2 & 2i\Omega\omega  \\
		-2i\Omega\omega & \omega^2 - k^2 v_A^2
	\end{array}
	\right) \dBp 
\]	
\begin{equation}
\hfill	= ikc \left(
	\begin{array}{cc}
		2\Omega & i\omega  \\
		-i\omega & \half \Omega
	\end{array}
	\right)\dEp \,. \qquad
	\label{eq:linearized_induction}
\end{equation}
The effects of finite conductivity are contained on the right hand 
side.  In ideal MHD, $ \dEp = 0 $, and the dispersion relation is 
obtained by setting the determinant of the matrix on the left to 
zero.  In the absence of rotation, this leads to the dispersion 
relation for Alfv\'en waves; with $ \Omega\neq 0$ the dispersion relation 
for the magnetorotational instability is obtained (cf. Balbus \& 
Hawley 1991).

The conductivity tensor relates $ \dEp $ to $ \dJp $, and hence 
via (\ref{eq:deltaJperp}) to another relationship between $ \dBp $ 
and $ \dEp $:
\begin{equation}
	\dEp = - \frac{ikc}{4\pi\sigma_{\perp}^2}\left(
	\begin{array}{cc}
		s\sigma_1 & -\sigma_2  \\
		\sigma_2 & s\sigma_1
	\end{array}
	\right) \dBp \,,
	\label{eq:dEp_conductivity}
\end{equation}
where
\begin{equation}
	s = \mathrm{sign}(B_z).
	\label{eq:s}
\end{equation}
Ideal MHD is a good approximation for long wavelength perturbations, 
when (\ref{eq:dEp_conductivity}) shows that $ |\dEp| $ is small.  
It breaks down when $ |\dEp| $ implied by 
(\ref{eq:dEp_conductivity}) becomes significant in the linearized 
induction equation (\ref{eq:linearized_induction}), that is when
$ k^2 c^2 / (4\pi \sigma_{\perp}) \ga |\omega| $.  This corresponds 
to $ kv_A \ga \omega_c $, where the critical frequency
\begin{equation}
	\omega_c = \frac{B^2\sigma_{\perp}}{\rho c^2}\,,
	\label{eq:crit_freq}
\end{equation}
is the generalisation of the ion-neutral collision frequency in the 
ambipolar diffusion case (Wardle \& Ng 1999).

The strength of the coupling between the magnetic field and the disc 
is determined by the parameter
\begin{equation}
	\chi = \frac{\omega_c}{\Omega} \,,
	\label{eq:eta}
\end{equation}
which is much larger or less than unity if the field is 
well or poorly coupled to the disc respectively. $\chi$ reduces
to the parameter $ \gamma\rho_i / \Omega $ that appears in 
the ambipolar diffusion case (Wardle \& K\"onigl 1993; Blaes \& Balbus 
1994).  If the diffusion is Ohmic, then $\chi = 
\Omega/\Omega_c$, where $\Omega_c = k^2 c^2 / 4\pi\sigma_2$ is the 
diffusion rate for waves with $k = \Omega/v_A$ (cf.  Jin 1996).

Substituting for $ \dEp $ in (\ref{eq:linearized_induction}) yields 
a matrix whose determinant must vanish if nonzero solutions for the 
perturbations are to exist.  The resulting dispersion relation is of the form 
\begin{equation}
	a (kv_A/\Omega)^4 + b (kv_A/\Omega)^2 + c = 0 \,
	\label{eq:dispersion_relation}
\end{equation}
where
\begin{equation}
	a = \chi^2 + \left( {\textstyle \frac{5}{2}}s\sigma_1
+ 2 \sigma_2\nu \right) \sigma_{\perp}^{-1}\chi + 1 + \nu^2 \,,
	\label{eq:acoeff}
\end{equation}
\begin{equation}
	b= (2\nu^2 - 3 ) \chi^2 + \left(2\sigma_2\nu - 
	{\textstyle \frac{3}{2}}s\sigma_1\right)(1 +\nu^2)\sigma_{\perp}^{-1}\chi \,,
	\label{eq:bcoeff}
\end{equation}
\begin{equation}
	c=\chi^2\nu^2(1+\nu^2) \,,
	\label{eq:ccoeff}
\end{equation}
and $ \nu=i\omega/\Omega $ (so that unstable modes have $ Re(\nu) > 0 $).
The discriminant is
\[	b^2-4ac = 
	\left[\left(9-16\nu^2\right) 
	\left( \chi+\half s\sigma_1\sigma_{\perp}^{-1}(1+\nu^2)\right) \right.
\]
\begin{equation}
    \left.
	- 12\sigma_2 \sigma_{\perp}^{-1}\nu (1+\nu^2)\right]
	\left[\chi+\half s\sigma_1 \sigma_{\perp}^{-1}(1+\nu^2)\right] \,.
	\label{eq:discriminant}
\end{equation}
$\sigma_{\parallel}$ does not appear in the dispersion relation and 
the linearized equations because $\k\cdot\B = 0$.  Thus the 
instability in an initially vertical magnetic field behaves 
identically in both the ambipolar diffusion ($\sigma_1 = 0$, 
$\sigma_{\parallel}=0$) and Ohmic diffusion ($\sigma_1 = 0$, 
$\sigma_{\parallel}=\sigma_2$) limits.

\section{Results}

The dispersion relation (\ref{eq:dispersion_relation}) is quadratic in 
$ k^2 $, and for real $ \nu $, $c$ is positive.  Thus there 
is at most one real, positive root to the dispersion relation.  
Typically the plots of $ \nu $ vs $ k $ resemble inverted quadratics, 
the growth rate increasing from zero at $ k=0 $, rising to a maximum 
(denoted by $ \nu_0\Omega $) at some wave number $ k_0 $, and decreasing to 
zero at a critical wave number $ k_c$.  However, I show below that for 
a small region of parameter space, $ k_c $ becomes infinite, and 
growth may occur over a wide range of wave numbers.

\subsection{Ideal MHD limit ($ \chi \rightarrow \infty $)}
Ideal MHD is recovered in the limit $ \omega_c \gg \Omega $, in 
which case the dispersion relation reduces to 
\begin{equation}
	(kv_A/\Omega)^4 + (2\nu^2 - 3) (kv_A/\Omega)^2 + \nu^2 (1 + \nu^2) = 0
	\label{eq:ideal_dispersion_relation}
\end{equation}
As shown by Balbus \& Hawley (1991), this yields growing modes for 
$ 0 \leq kv_A/\Omega \leq \sqrt{3} $, with the maximum growth rate
$ \nu_0 = \frac{3}{4} $ occuring 
when the discriminant of eq. (\ref{eq:ideal_dispersion_relation}) 
vanishes, that is at $ k_0v_A/\Omega = \sqrt{\frac{15}{16}} $.

\subsection{Ambipolar or Ohmic diffusion limit ($\sigma_1 = 0$)}
\begin{figure} %\subsubsection{fig:nu_vs_k_a}
\centerline{\epsfxsize=8cm\epsfbox{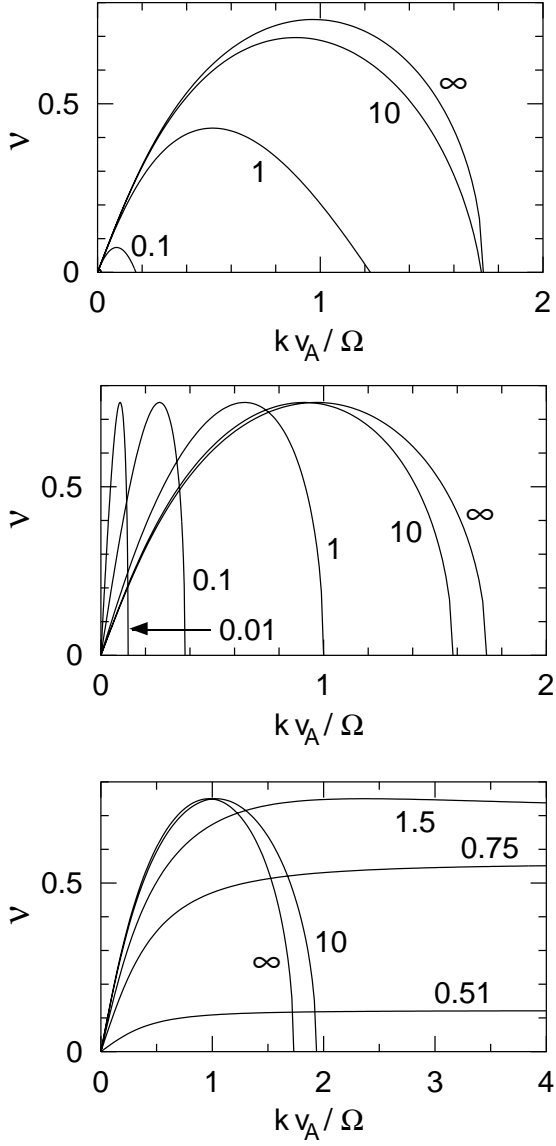}}
\caption{The growth rate of the Balbus-Hawley instability as a 
function of wavenumber for different choices of the Hall and Pedersen 
conductivities $\sigma_1$ and $\sigma_2$ (see text).  \emph{Upper:} 
the ambipolar or Ohmic diffusion limits ($\sigma_1 = 0$); \emph{middle:} the 
Hall limit ($\sigma_2 = 0$) with $\sigma_1 B_z>0$; and \emph{lower:} 
the Hall limit for $\sigma_1 B_z<0$.  The curves in each panel are 
labelled by $\chi$, the ratio of the coupling frequency to the disc 
angular frequency: ideal MHD is recovered for 
$\chi\rightarrow\infty$, and the magnetic field is poorly coupled to 
the disc for $\chi < 1 $.  Note that the scale on the horizontal axis 
differs between the lower and upper two plots -- the ideal MHD curve 
labelled ``$\infty$'' appears in each plot.}
\label{fig:nu_vs_k_a}
\end{figure}

In this case  (c.f. Blaes \& Balbus 1994) growth occurs up to a wave number
\begin{equation}
	k_c v_A/\Omega= \sqrt{3}
	\left(1+ \chi^{-2} \right)^{1/2} \, 
	\label{eq:kcrit_ambdiff}
\end{equation}
and the maximimum growth rate, again identified by requiring 
(\ref{eq:discriminant}) to vanish, is the positive real solution to 
\begin{equation}
	12\nu_0^3 + 16\chi \nu_0^2 - 9\chi + 12 = 0 \,.
	\label{eq:nu_max_ambdiff}
\end{equation}
For $ \chi \gg 1 $, $ \nu_0 $ is close to the ideal MHD value of $ 
\frac{3}{4} $, declining substantially for $ \chi \la 1 $.  For $ \chi 
\ll 1 $, $ \nu_0\approx \frac{3}{4} \chi$, and $ k_0 v_A/\Omega 
\approx \sqrt{\frac{3}{4}}\chi $.  This is illustrated in the top 
panel of Fig.  \ref{fig:nu_vs_k_a} which shows the growth rate as a 
function of wave number for $ \chi = \infty $, 10, 1, and 0.1.  There 
is not much change from the ideal limit until $ \chi < 10 $, when the 
growth rate and characteristic wave number decline rapidly as $\chi$ 
is reduced.

\subsection{Hall limit ($\sigma_2 = 0$) with $\sigma_1B_z >0$}
The departure from the ideal limit changes markedly when the Hall term 
is introduced.  In the absence of ambipolar diffusion, with 
$\sigma_1B_z >0$, one finds growth for wavenumbers between 0 and 
\begin{equation}
	k_c v_A/\Omega = 
	\sqrt{3}\left(1+2\chi^{-1}\right)^{-1/2}\,,
	\label{eq:kcrit_0}
\end{equation}
and a maximum growth rate \emph{unchanged} from the ideal limit, i.e. $ 
\nu_0 = \frac{3}{4} \, $!   The corresponding wave number, however,
decreases as the geometric mean of $ \Omega $ and $ \omega_c $ 
as $\chi \rightarrow 0 $:
\begin{equation}
	k_0 v_A /\Omega = 
	\sqrt{{\textstyle \frac{15}{16}}} 
	\left(1+{\textstyle \frac{5}{4}}\chi^{-1} \right)^{-1/2} \,.
	\label{eq:kmax_0}
\end{equation}
This behaviour is demonstrated in the middle panel of Fig. \ref{fig:nu_vs_k_a}.

\subsection{Hall limit ($\sigma_2 = 0$) with $\sigma_1B_z <0$}

If the other Hall-dominated limit is taken, the behaviour is again 
different (see lower panel of Fig. \ref{fig:nu_vs_k_a}).  
For $ \chi > 2 $, the 
range of wave numbers for which growth occurs is finite, with 
\begin{equation}
	k_c v_A /\Omega = \sqrt{3}\left(1-2\chi^{-1}\right)^{-1/2} \,.
	\label{eq:kcrit_180}
\end{equation}
$k_c$ becomes infinite once $ \chi $ reaches 2, and for $ 
\frac{1}{2} < \chi < 2 $, all wave numbers grow.  There is a 
fastest-growing mode for $ \chi > \frac{5}{4} $, with $ \nu_0 = 
\frac{3}{4} $ and
\begin{equation}
	k_0 v_A /\Omega= \sqrt{{\textstyle \frac{15}{16}}}\left(1-{\textstyle 
	\frac{5}{4}}\chi^{-1}\right)^{-1/2}
	\label{eq:kmax_180}
\end{equation}
(cf. the $\chi=1.5$ case in the lower panel of Fig. \ref{fig:nu_vs_k_a}).
Again, this becomes infinite for $ \chi = \frac{5}{4} $
For $ \frac{1}{2} < \chi < \frac{5}{4} $, the growth rate is a 
monotonically increasing function of $ k $, asymptotically 
approaching
\begin{equation}
	\nu_0 = (\chi-\half)^{1/2}(2-\chi)^{1/2} 
	\label{eq:nmax_180}
\end{equation}
This limit decreases to zero for $ \chi=\frac{1}{2} $ (cf.  the 
$\chi=0.75$ and 0.51 curves in the lower panel of Fig.  
\ref{fig:nu_vs_k_a}), at which point there are no longer any growing 
modes.

\subsection{The general case}

When $ \sigma_1 B_z > 0 $, there are growing modes with wave numbers 
between 0 and
\begin{equation}
	k_c v_A / \Omega = \left(\frac{3\chi(\chi+\half s \sigma_1 
	\sigma_{\perp}^{-1})}{\chi^2 + {\textstyle \frac{5}{2}} \chi s 
	\sigma_1 \sigma_{\perp}^{-1} + 1}\right)^{1/2} .
	\label{eq:kcrit_general}
\end{equation}
The maximum growth rate satisfies
\begin{equation}
	\left(\frac{12\sigma_2 \sigma_{\perp}^{-1}\,\nu_0}{9-16\nu_0^2}
	-\half s \sigma_1 \sigma_{\perp}^{-1} \right) 
	(1+\nu_0^2) = \chi \,.
	\label{eq:nmax_general}
\end{equation}
\begin{figure} %\subsubsection{fig:nu_vs_k_b}
\centerline{\epsfxsize=8cm\epsfbox{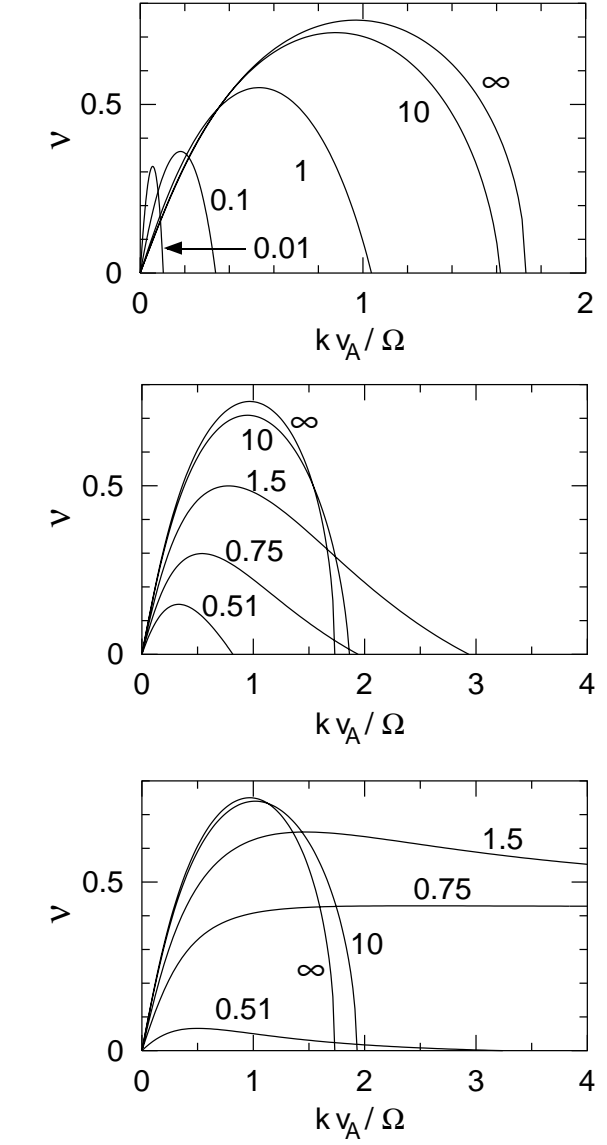}}
\caption{As for Fig.  \ref{fig:nu_vs_k_a}, but for intermediate cases.  
\emph{Upper panel:} $\sigma_1 B_z > 0$ and $|\sigma_1| = \sigma_2 $; 
\emph{middle panel:} $\sigma_1 B_z < 0$ and $|\sigma_1| = \sigma_2 $; 
and \emph{lower panel:} $\sigma_1 B_z < 0$ with $|\sigma_1| = 6 
\sigma_2 $.}
\label{fig:nu_vs_k_b}
\end{figure}
This behavior is illustrated in the top panel of Fig.  
\ref{fig:nu_vs_k_b}, which shows the wave number dependence of the 
growth rate when $|\sigma_1| = \sigma_2$.  As $ \chi $ is reduced, the 
maximum growth rate and associated wave number decrease.  However, in 
the limit $ \chi\rightarrow 0 $, the growth rate remains finite, with
\begin{equation}
	\nu_0 \approx \frac{3}{4} \,
	 \frac{|\sigma_1|}{\sigma_{\perp}+\sigma_2}\, \,.
	\label{eq:max_growth_small_eta}
\end{equation}
The corresponding wave number, however, decreases:
\begin{equation}
	k_0 v_A / \Omega \approx 
	\left(\frac{3\chi}{4}\frac{|\sigma_1|}{\sigma_{\perp}+\sigma_2}\right)^{1/2}.
	\label{eq:max_k_small_eta}
\end{equation}

When $ \sigma_1 B_z < 0 $, growing modes exist only for $ 
\chi > \half |\sigma_1|/\sigma_{\perp} $. 
If $ |\sigma_1|/\sigma_{\perp} > \frac{4}{5} $,  then 
eqs (\ref{eq:kcrit_general}) and (\ref{eq:nmax_general}) hold; otherwise
there is a restricted range of $ \chi $ for which all wave numbers 
grow, determined by the inequality
\begin{equation}
	\left( \chi -  \frac{5|\sigma_1|}{4\sigma_{\perp}} \right)^2 < 
	\left(\frac{5|\sigma_1|}{4\sigma_{\perp}}\right)^2 - 1 \,.
	\label{eq:eta_range}
\end{equation}
The middle 
panel of Fig. \ref{fig:nu_vs_k_b}, shows the behaviour as $ \chi $ is reduced
for $ |\sigma_1|=\sigma_2 $.  In this case the range of growing modes 
is always finite.  The 
lower panel, for $ |\sigma_1| = 6 \sigma_2 $ demonstrates the existence of a
small range of $ \chi $ around 0.75 over which all wave numbers grow.

\begin{figure} %\subsubsection{fig:max_growth_rate}
\centerline{\epsfxsize=8cm\epsfbox{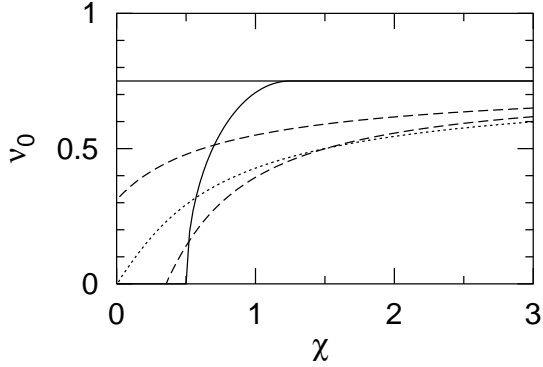}}
\caption{Maximum growth rate (in units of the Keplerian frequency 
$\Omega$) as a function of degree of coupling between the magnetic 
field and neutral gas, $\chi$, for different choices of the Hall and 
Pederson conductivities $\sigma_1$ and $\sigma_2$.  \emph{Dotted 
curve:} the ambipolar and Ohmic diffusion limits ($\sigma_1 = 0$); 
\emph{solid:} the Hall limit ($\sigma_2 = 0$); \emph{dashed:} the case 
$|\sigma_1|=\sigma_2$.  The upper and lower members of the pairs of 
dashed or solid curves correspond to $\sigma_1 B_z > 0$ or $\sigma_1 
B_z < 0$ respectively.}
 \label{fig:max_growth_rate}
\end{figure}

\begin{figure} %\subsubsection{fig:max_growth_rate_contours}
\centerline{\epsfxsize=8cm\epsfbox{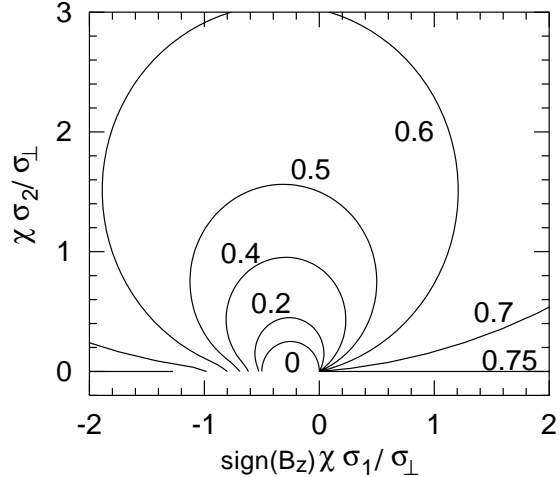}}
\caption{A polar plot of the maximum growth rate of the Balbus-Hawley 
instability.  The Cartesian cordinates of a point on the plane 
corresponds to $(\mathrm{sign}(B_z)\sigma_1,\sigma_2)$ suitably 
normalised.  The polar radius is $\chi$, the parameter controlling the 
degree of magnetic coupling in the disc.  The contours are labelled by 
the maximum growth rate in units of $\Omega^{-1}$.  }
\label{fig:max_growth_rate_contours}
\end{figure}

These results are summarised in Figures \ref{fig:max_growth_rate} and 
\ref{fig:max_growth_rate_contours}.  Fig. \ref{fig:max_growth_rate} 
shows the maximum growth rate as a function of $ \chi $ for different 
choices of $\sigma_1 $ and $\sigma_2$.  In the ambipolar and Ohmic 
diffusion limits ($\sigma_1 =0$), the maximum growth rate decreases to 
zero as $ \chi \rightarrow 0 $.  When the Hall term is included, the 
growth rate drops to zero at $ \chi = \half |\sigma_1|/\sigma_{\perp} 
$ for $ \sigma_1 B_z < 0 $, or remains finite at $ 
\frac{3}{4}|\sigma_1| / (\sigma_{\perp}+\sigma_2) $ for $ \sigma_1 B_z 
> 0 $.
This behaviour is displayed more fully in Fig.  
\ref{fig:max_growth_rate_contours}, which shows contours of fixed 
growth rate in a polar plot.  The ambipolar and Ohmic diffusion limits 
apply along the locus $\sigma_1 = 0$ running vertically through the 
origin.  It is clear that the growth rate in the weak coupling limit 
($\chi \la 1$) is sensitive to the form of the conductivity tensor, 
and that the behaviour in the two diffusion limits is not generally 
representative.

\section{A two-fluid model for the Hall limit}
\label{sec:two-fluid}
It is surprising at first glance that the Hall conductivity allows the 
Balbus-Hawley instability to grow at a rate of order $ \Omega $ in the 
formal limit of zero coupling.  Here I address the physical 
explanation for this result, for simplicity focussing on the ``pure'' 
Hall case ($\sigma_2=0$).  In the ambipolar 
diffusion limit, intuition is aided by a model two-fluid system: an 
ionized (but electrically-neutral) fluid into which the magnetic field 
is frozen; and a neutral fluid which is indirectly coupled to the 
magnetic field via collisions of the charged speicies with the 
neutrals.  There is an analogous model in the Hall limit $|\sigma_1| 
\gg \sigma_2$.

In this model, the ionized component of the fluid can be regarded as 
consisting of two species, one of which does not suffer significant 
collisions with the neutrals, the other being so strongly tied to the 
neutrals by collisions that its drift velocity is negligible.  For 
convenience, I shall refer to the former as ``electrons'' (subscript 
$e$) the latter as ``ions'' (subscript $i$); although the roles of 
these species could be played by other species, e.g.  ions and grains, 
under suitable conditions.

The Hall parameters of the two species satisfy $ |\beta_i| \ll 1 \ll 
|\beta_e| $.  Because the ions are strongly tied to the neutrals by 
collisions and are responsible for transmitting the electromagnetic 
stresses to the fluid, one may regard the ions and neutrals as a 
single fluid with the charge density of the ions and the inertia (and 
thermal pressure, when relevant) of the neutrals.  This fluid is 
permeated by the electrons, into which the magnetic field is frozen.
 
In this limit the components of the conductivity perpendicular 
to the magnetic field are
\begin{equation}
	\sigma_1 \approx \frac{ec n_i Z_i}{B}
	\label{eq:sigma1_Hall}
\end{equation}
and
\begin{equation}
	\sigma_2 \approx \left(|\beta_i|+|\beta_e|^{-1}\right) |\sigma_1|
	\label{eq:sigma2_Hall}
\end{equation}
(where I have used $ n_iZ_i = - n_eZ_e $ in eq.  
\ref{eq:sigma1_Hall}).  The current is dominated by the electrons, 
which drift perpendicular to the magnetic and electric field so that 
the net Lorentz force on them is zero -- collisions with the neutrals 
play a negligible role in their equation of motion.  The drift of the 
ions with respect to the neutrals, however, is strongly inhibited by 
collisions, and is determined by the balance between the electric 
field and the drag associated with the neutrals.  Thus
\begin{equation}
	\J\approx n_eZ_e e\v_e \approx \sigma_1 \Bh\cross\Epe \,.
	\label{eq:J_Hall}
\end{equation}
The electromagnetic stresses on the fluid are communicated largely by the 
collisions with the ions:
\begin{equation}
	\frac{\J\cross\B}{c}\approx \sigma_1 \Epe \approx 
	\gamma_i\rho\rho_i\v_i
	\label{eq:drag_Hall}
\end{equation}
Because the ion-neutral drift speed $\v_i$ is small, the associated 
dissipation is small: $\J\cdot\Epe \approx \sigma_2 E_{\perp}'^2$,
tending to zero in the limit $ \sigma_2 \rightarrow 0 $.  
With this model fluid, one can use the linearized equations 
(\ref{eq:deltaJperp})--(\ref{eq:dEp_conductivity}) to understand the 
growth of the instability in the Hall limit as $\chi \rightarrow 0$.  

First consider the wave modes supported by the fluid in the absence of 
rotation.  At small wavenumbers (long wavelengths) the fluid supports 
the Alfv\'en waves of ideal MHD, while at wavenumbers far in excess of 
$ \omega_c / v_A $, there are two circularly-polarized modes (Wardle 
\& Ng 1999).  In one, the electrons and magnetic field are relatively 
unperturbed, and the ion-neutral fluid executes gyrations about the 
field direction at the frequency $\omega \approx \omega_c $.  In the 
other, the ion-neutral fluid is static and the electrons and magnetic 
field twist about the unperturbed field direction with the opposite 
sense of rotation.  The frequency of this mode scales quadratically 
with wavenumber, as the diffusive term $\curl \E'$ dominates $\curl 
(\v\cross\B)$ in the induction equation: $\omega \approx 
(kv_A)^2/\omega_c $.  The sense of rotation of either mode is 
determined by the sign of $\sigma_1 B_z$.

Finally, consider the effect of rotation, which attempts to enforce 
epicyclic motion of the ions and neutrals with frequency $\Omega$ 
through the Coriolis and angular terms in the momentum equation.  For 
$\chi \ll 1$, $\Omega \gg \omega_c $, so the rotation can only couple 
effectively to the second mode for wave numbers $k v_A \approx 
\sqrt{\omega_c \Omega}$.  The generation of toroidal field from 
poloidal field by the shear in the disc (see eq.  \ref{eq:induction}) 
acts as a forcing term for this circularly polarized mode, which grows 
or decays depending on whether the shear assists or opposes the 
rotation of the magnetic field, i.e.  depending on the sign of 
$\sigma_1 B_z$.

\section{Discussion}

The Hall conductivity modifies the growth of the Balbus-Hawley 
instability in weakly ionized discs when the magnetic field is not 
well-coupled to the neutral matter.  The effect depends on the 
relative signs of $ \sigma_1 $ and the initially vertical field $B_z$.  
For $ \sigma_1 B_z >0 $ the growth rate remains of order the Keplerian 
frequency $\Omega$ as the coupling is reduced, and the wavelength of 
the fastest growing mode scales as $ \chi^{1/2} $.  On the other hand, 
if $ \sigma_1 B_z < 0$, the mode is damped for $\chi \la \half$.  The 
dependence on the initial field being parallel or antiparallel to the 
disc rotation axis arises ultimately from the intrinsic handedness 
introduced into the medium by the difference in the drifts of positive 
and negative species.  In the limit $ \Omega \rightarrow 0 $, the 
unstable mode corresponds to a standing, circularly-polarized Alfv\'en 
wave; left and right-circularly polarized Alfv\'en waves propagate 
differently according to the sign of $ \sigma_1 B_z $ (Wardle \& Ng 
1999).

This is to be contrasted with the ambipolar and Ohmic diffusion limits 
(Blaes \& Balbus 1994; Jin 1996), where the growth rate is independent 
of the sign of $B_z$, declining steadily while the wavelength 
increases linearly as the coupling parameter is reduced.  Blaes and 
Balbus found that the growth rate increases again for $\chi \la \rho_i 
/ \rho$, when the coupling becomes so poor that the momentum-exchange 
time scale for the ions with the neutrals\footnote{As distinct from 
the momentum exchange time scale for the neutrals with the ions, 
$(\gamma\rho_i)^{-1}$ that appears in the definition of $\chi$ in the 
ambipolar diffusion limit.}, $(\gamma_i \rho)^{-1}$ is longer than the 
Keplerian time scale $\Omega^{-1}$.  Then the instability develops in 
the ionized component and magnetic field without affecting the 
neutrals.  This behaviour does not occur here because I neglected the 
inertia of the charged species (see Section ) -- the formulation in 
Section \ref{sec:formulation} is valid only for frequencies well below 
$\gamma_i \rho$, and thus implicitly assumes that $\Omega \ll \gamma_i 
\rho$, i.e.  $\chi \gg \rho_i / \rho$.  The ionization fraction in 
protostellar discs is so low that this approximation is very good 
indeed, and if the coupling were this poor any dynamically important 
magnetic field would remove itself (and the ions) from the disc on the 
Alfv\'en crossing time in the ionized fluid.  The ionized species 
cannot, therefore, decouple from the neutrals in any practical sense.

For a favourable alignment between the magnetic field and rotation 
axis (i.e.  $\sigma_1B_z>0$), the growth rate in the weak-coupling 
limit remains within a factor of a few of $ \Omega $.  The practical 
limit on the coupling is determined by the wavelength as the coupling 
gets weak, as this must be less than the scale height $ h $ of the 
disc.  If $ c_s $ is the isothermal sound speed, so that $ h = c_s 
/\Omega $, the wavelength of the fastest-growing mode becomes of order 
$ h $ when the coupling parameter $\chi$ reaches $v_A^2/c_s^2$.  The 
corresponding limit in the ambipolar diffusion case is $v_A/c_s$ which 
would produce a growth rate of order $(v_A/c_s)\Omega$.

A study of the nonlinear development of the instability requires a
numerical simulation.  For now, I note that the strength of the 
magnetic field affects the conductivity through the Hall parameters, 
with a larger field increasing the degree of flux-freezing in the 
ionized component.  Thus there is the possibility of the instability 
beginning in the Hall regime, with the field strength growing to the 
point that the fluid enters the ambipolar diffusion regime.

These results are applicable to protostellar discs, where collisions 
are sufficient to decouple ions and grains from the magnetic field 
(Wardle \& K\"onigl 1993).  Indeed, the Hall conductivity is of the 
same order as the ambipolar diffusion term over a broad range of 
parameters (Wardle \& Ng 1999).  Further the coupling parameter is 
thought to be small near the midplane in protostellar discs, 
particularly at radii around one AU where the temperature is low 
enough that thermal ionization is not yet effective, and the surface 
density is high enough to shield the material from cosmic rays 
(Hayashi 1981; Gammie 1996; Wardle 1997).  The magnitude and relative 
size of the components of the conductivity tensor vary strongly with 
height in the disc as the neutral density decreases near the disc 
surface, and the ionization level increases because of exposure to 
cosmic rays (Gammie 1996; Wardle 1997) and 
X-rays from the central star (Glassgold et al 1997).  In particular, 
although the instability is not very efficient in transporting angular 
momentum when the coupling is weak, the column density within this 
portion of the disc is relatively large, so that the net angular 
momentum transport can be comparable to that in the layers closer to 
the surface.

The dependence on the sign of the initial field raises the intriguing 
possibility that the accretion process, and the formation of 
magnetically-driven disc winds are dependent on the sign of the 
magnetic field.  This issue is linked to the collapse of cloud cores 
to form protostars and discs, which presumably brings in some fraction 
of the interstellar field to be the initial disc field.  While the 
calculations of core collapse to date are instructive (e.g.  Fiedler 
\& Mouschovias 1993; Basu 1997; Contopoulos, Ciolek, \& Konigl 1998; 
Li 1998; Ciolek \& K\"onigl 1998), none include the Hall effects, 
which are also important during this phase (Wardle \& Ng 1999).  
Angular momentum transport during collapse will also depend on whether 
the field is parallel or antiparallel to the core's angular momentum 
vector, and one might imagine that this leads to a preferential 
orientation within discs, but this is unclear as the sign of $ 
\sigma_1 $ may change between the collapse and disc phases as grains, 
then ions become successively decoupled from the magnetic field with 
increasing neutral density.

\section{Summary}
\label{sec:summary}
This paper examined the Balbus Hawley instability of a vertical 
magnetic field in a weakly-ionized, thin Keplerian disc.  The vertical 
stratification was ignored, and radial and azimuthal variations were 
neglected.  The resulting linearized equations are valid for 
wavelengths small compared to the disc scale height.  I examined the 
dispersion relation.  The conductivity tensor contributes two 
parameters: \emph{(i)} a coupling parameter $ \chi $, the ratio of the 
frequency at which the finite conductivity becomes important to the 
Keplerian angular frequency; and \emph{(ii)} the ratio $ 
\mathrm{sign}(B_z) \sigma_1/ \sigma_{\perp}$, which reflects the 
contributions of the Hall and Pedersen conductivities, $ \sigma_1 $, $ 
\sigma_2 $ respectively, to the total conductivity perpendicular to 
the magnetic field $\sigma_{\perp} = (\sigma_1^2+\sigma_2^2)^{1/2}$.

I found the following:
\begin{enumerate}
	\item The Hall conductivity substantively modifies the behaviour 
	of the instability once finite conductivity effects become 
	important ($\chi\la 1)$.  If the rotation axis of the disc defines 
	the $ z $ direction, the departure from the ambipolar diffusion 
	limit depends on the sign of $ \sigma_1 B_z $.
	
	\item For $ \sigma_1 B_z > 0 $, growth occurs over a finite range 
	of wavenumbers.  For poor coupling ($ \chi \ll 1 
	$), the maximum growth rate remains finite, tending to a value $ 
	\frac{3}{4}|\sigma_1| / (\sigma_{\perp}+\sigma_2) \, \Omega$ while 
	the associated wave number scales as $ \sqrt{\chi} $.  This should 
	be contrasted with the ambipolar diffusion case (Blaes \& Balbus 
	1994) for which the growth rate and wave number scale linearly in 
	$\chi$.

	\item For $\sigma_1 B_z < 0$, on the other hand, there is no 
	instability when the coupling is poor: growth requires $\chi > 
	\half|\sigma_1|/ \sigma_{\perp}$.  There is generally a limited 
	range of wave numbers, over which 
	growth occurs.  However, if $ |\sigma_1|/ \sigma_{\perp} < 
	\frac{4}{5} $, there is a limited range of $\chi$ around the value 
	$ |\sigma_1|/ \sigma_{\perp} $ for which \emph{all} wave numbers 
	grow.

	\item Although the instability has a formal growth rate of order $ 
	\Omega $ in the limit of zero coupling when $\sigma_1 B_z > 0$, 
	the wavelength is limited by the disc scale height, and therefore 
	growth will only occur for $ \chi \ga v_A^2/c_s^2 $.  
	Note that this constraint is far less severe than in the ambipolar 
	diffusion limit, which requires $\chi \ga v_A/c_s$.

\end{enumerate}
Finally, I stress that these results indicate the 
qualitative changes to dynamical behaviour to be expected when 
ambipolar diffusion breaks down during the collapse of cloud cores to 
form protostars and during the subsequent evolution of their attendant 
discs.

I thank Charles Gammie and the referee for comments on the manuscript.
The Special Research Centre for Theoretical Astrophysics is funded 
under the Special Research Centres programme of the Australian 
Research Council. 

%\section{References}

\bsp
\label{lastpage}
\end{document}